\documentclass{pasj00}
\draft

\begin{document}
\SetRunningHead{Kohno et al.}{Enhanced HCN Emission in the Seyfert Galaxy NGC 1097}
\Received{}
\Accepted{}

\title{Enhanced HCN(1--0) Emission in the Type-1 Seyfert Galaxy NGC 1097}

\author{ 
Kotaro \textsc{Kohno},\altaffilmark{1}
Sumio \textsc{Ishizuki},\altaffilmark{2}
Satoki \textsc{Matsushita},\altaffilmark{3} \\
Baltasar \textsc{Vila-Vilar\'o},\altaffilmark{4}
and
Ryohei \textsc{Kawabe}\altaffilmark{2}
}

\altaffiltext{1}{\it Institute of Astronomy, The University of Tokyo, Osawa, Mitaka, Tokyo, 181-8588}
\email{kkohno@ioa.s.u-tokyo.ac.jp}
\altaffiltext{2}{\it National Astronomical Observatory, Osawa, Mitaka, Tokyo, 181-8588}
\altaffiltext{3}{\it Submillimeter Array, Harvard-Smithsonian Center for Astrophysics,\\ P.O. Box 824, Hilo, HI 96721-
0824, U.S.A.}
\altaffiltext{4}{\it Steward Observatory, The University of Arizona, Tucson, AZ 85721, U.S.A.}

%
%
%


\maketitle

\begin{abstract}
The central kpc region of the low luminosity type-1 Seyfert galaxy
NGC 1097 has been observed in the CO(1--0) and HCN(1--0) lines
with the Nobeyama Millimeter Array and the NRO 45 m telescope.
We find a striking enhancement of the HCN emission 
toward the active nucleus of NGC 1097;
a large fraction of the CO emission comes from the circumnuclear
starburst ring ($r \sim 10''$ or 700 pc at $D = 14.5$ Mpc),
whereas the HCN emission is dominated by a strong unresolved peak 
at the nucleus.
The HCN-to-CO integrated intensity ratio 
in brightness temperature scale, $R_{\rm HCN/CO}$,
is about 0.11 in the aperture of the 45 m telescope beams 
($r<510$ pc for CO and $r<640$ pc for HCN),
but it reaches 0.34 in the smaller aperture 
(within the central $r< 350 \times 150$ pc region).
These CO and HCN properties in NGC 1097 are similar to those in the type-2 Seyfert
galaxies NGC 1068 and NGC 5194 (M 51).

\end{abstract}

\section{Introduction}

The dense molecular medium plays various roles
in the vicinity of active galactic nuclei (AGNs).
The presence of dense and dusty interstellar matter (ISM),
which obscures the broad line regions in AGNs,
is inevitable at $<1$ pc - a few 10 pc scales
according to the proposed unified model of Seyfert galaxies 
(e.g., \cite{ant93}).
These circumnuclear dense ISM
could be a reservoir of fuel for the active nuclei,
and also be a site of massive star formation.
In fact, strong HCN(1--0) emission, which requires
dense ($n_{\rm H_2} > 10^4$ cm$^{-3}$) environments
for its collisional excitation, has been detected
in the prototypical type-2 Seyfert NGC 1068
(Jackson et al.\ 1993; Tacconi et al.\ 1994;
\cite{hb95}) and the low-luminosity
type-2 Seyfert NGC 5194 (Kohno et al.\ 1996).
The HCN(1--0) to CO(1--0) integrated intensity ratios in brightness
temperature scale, $R_{\rm HCN/CO}$,
are enhanced up to about 0.4 -- 0.6
in these Seyfert nuclei, and the kinematics
of the HCN line imply that this dense molecular medium
can be outer envelopes of the predicted obscuring material.
Nevertheless, our knowledge on the nature of 
circumnuclear dense molecular matter in Seyfert
nuclei has still been very limited.
This is because there are only a few high resolution
(i.e., a few 100 pc resolution or less) and sensitive
observations of dense molecular gas in Seyfert galaxies,
although several surveys of dense molecular gas
with low resolution instruments have been conducted 
to address the global quantities of dense molecular gas
in the AGN hosts (\cite{ps98,cur00}).

In this paper, we report detection of
the enhanced HCN emission toward the nucleus
of the low luminosity type-1 Seyfert galaxy NGC 1097 
using the Nobeyama Millimeter Array (NMA) and NRO 45 m telescope.
NGC 1097 is a nearby ($D$ = 14.5 Mpc, \cite{tul88})
barred spiral galaxy classified as SB(s)b 
(de Vaucouleurs et al.\ 1991)
showing double-peaked broad H$\alpha$ emission with time variability 
at the nucleus (Storchi-Bergmann et al.\ 1997). 
The detection of a hard X-ray source at the nucleus (\cite{iyo96})
also supports the presence of a genuine active nucleus, 
though it is rather a low luminosity one 
($L_{\rm 2 - 10 keV} = 3.7 \times 10^{40}$ erg s$^{-1}$
at $D$ = 14.5 Mpc).  
Two pairs of huge ($\sim$ a few 10 kpc scale) optical jets have been reported
yet their nature is still unclear (e.g., \cite{weh97}).
In the circumnuclear region of NGC 1097, there is a well-known 
starburst ring with a radius of $\sim 10''$ or 700 pc,
which is luminous at various wavelengths including radio (\cite{hum87}),
mid-infrared (\cite{tel93}), optical (Barth et al.\ 1995; Quillen et al.\ 1995;
Storchi-Bergmann et al.\ 1996), near-infrared (Kotilainen et al.\ 2000),
and soft X-ray (P\'erez-Olea, Colina 1996).
The star formation rate (SFR) in the ring is very high; 
about 5 $M_\odot$ yr$^{-1}$ from the extinction corrected
H$\alpha$ luminosity (\cite{hum87}).
The host of NGC 1097 seems to have nested bars (\cite{sha93}).
CO emission in the center of NGC 1097 has been mapped
with the NRO 45 m telescope (\cite{gnc88})
and a molecular ring associated with the starburst ring
has been detected, although angular resolution is insufficient
to fully resolve the structure. 
Low resolution ($\sim 1'$) single dish observations 
of the CO and HCN lines in NGC 1097 are also reported (\cite{hb93}).

\section{Observations and Data Reduction}

The NMA observations of CO(1--0) and HCN(1--0) were made
during the period from 1999 November to 2000 March.
The NMA consists of six 10 m antennas 
equipped with tunerless DSB SIS receivers. 
Two antenna configurations (C and D) were used.
The backend was the Ultra Wide-Band Correlator
(UWBC; \cite{oku00}). It was configured
to cover 512 MHz (1330 km s$^{-1}$) 
at 2 MHz resolution for CO,
and to cover 1024 MHz (3480 km s$^{-1}$) 
at 8 MHz resolution for HCN.
The visibility calibrator 0202-172 ($\sim$ 1.1 Jy) 
was observed every $\sim$ 20 minutes, and 3C454.3 was observed
to determine the passband.
The resultant CO (HCN) cube has a velocity resolution
of 15.6 km s$^{-1}$ (27.2 km s$^{-1}$) and a typical rms noise 
of 29 mJy beam$^{-1}$ (6 mJy beam$^{-1}$) for each channel.

The 45 m telescope observations were performed
in 1998 January. CO(1--0) and HCN(1--0) were
observed simultaneously using two SIS receivers
with side-band rejection filters (S100 and S80).
The sky emission was removed by position switching
at an offset of $3'$ in azimuth from the center.
The pointing accuracy was checked every 1 hour 
using a SiO maser source.  
The spectra were acquired with the 250 MHz wide
(650 km s$^{-1}$ for CO and 850 km s$^{-1}$ for HCN)
acousto-optical spectrometers. 
Because the velocity width of emission is
very large ($\sim 500$ km s$^{-1}$) compared with
the spectrometer  bandwidth, each spectrum was inspected carefully. 
After subtracting linear baselines,
adjacent channels were smoothed to 10 km s$^{-1}$
for CO and 20 km s$^{-1}$ for HCN.
The rms noise level was 16 mK for CO
and 3.5 mK for HCN on a $T_{\rm A}^*$ scale.

\section{Results}

We display the CO and HCN velocity integrated intensity maps
taken with the NMA in figure 1, together with an optical image
and the intensity-weighted mean velocity map of CO emission.
Strong enhancement of the HCN emission toward the nucleus of NGC 1097
is immediately evident; the CO image shows three
major peaks, i.e., the central peak and ``twin-peaks'' 
where dust lanes are connected to the circumnuclear starburst ring. 
On the other hand, the most conspicuous emission comes from the nucleus
in the HCN map. 

The azimuthally averaged radial distributions of the CO and HCN intensities 
and the HCN/CO ratios are shown in figure \ref{fig:radialdist}.
Again it is confirmed that a considerable fraction of CO emission
comes from the ring, whereas the HCN is dominated by the nuclear peak.
The resultant HCN/CO ratio is very high; it reaches 0.34 
within the central $r<5''\times2''$ or $350 \times 150$ pc region.
Note that the synthesized beams of these NMA observations
are elongated along the North to South direction
due to the low declination of NGC 1097,
and the observed $R_{\rm HCN/CO}$
is highly smeared with the observing beam.
We expect that the line ratio toward the nucleus
would increas further with higher angular resolutions.
Additional NMA observations
with the most extended configuration are in progress, 
and will be presented in forthcoming papers.

The total CO and HCN fluxes within the NMA field of views were about
690 $\pm$ 20 Jy km s$^{-1}$ and 67 $\pm$ 2.9 Jy km s$^{-1}$,
respectively.
The total CO flux corresponds to the molecular gas mass
of 1.4 $\times$ $10^9$ $M_\odot$ (including He and heavier
elements) if a Galactic $N_{\rm H_2}/I_{\rm CO}$
conversion factor, $X_{\rm CO} = 1.8 \times 10^{20}$
cm$^{-2}$ (K km s$^{-1}$)$^{-1}$ (\cite{dame01}), is applied
just for comparison purposes.

We show the CO and HCN spectra toward the center of NGC 1097
taken with the 45 m telescope in figure \ref{fig:45spectra}. 
Integrated intensities of
CO and HCN were $96\pm2.8$ and $11\pm0.71$ K km s$^{-1}$ 
on a $T_{\rm MB} = T_{\rm A}^*/\eta_{\rm MB}$ scale.
The obtained CO spectrum roughly agrees with the previous observations
(\cite{gnc88}), but the quality of our data is greatly improved.
The HCN-to-CO integrated intensity ratio was 0.11 
at the 45 m telescope observing beams
($r<510$ pc for CO and $r<640$ pc for HCN).

We compared the NMA fluxes of CO and HCN with 45 m telescope fluxes
by convolving the NMA cubes to the same beam sizes as the 45 m observations.
We find that the NMA CO flux corresponds to about 0.88 of the 45 m CO flux,
and that the NMA HCN flux is mostly the same as the 45 m HCN flux within the errors.
Thus, most of the single dish fluxes are recovered
by our interferometric observations.

\section{Discussion}

\subsection{Nature of Dense Molecular Gas toward the Seyfert Nucleus}

We have detected enhanced HCN emission toward the nucleus
of the type-1 Seyfert galaxy NGC 1097. 
The $R_{\rm HCN/CO}$ is about 0.11 
in the aperture of the 45 m telescope beams
($r<$ \timeform{7".5} or 510 pc for CO
and $r<$ \timeform{9".5} or 640 pc for HCN),
but it reaches 0.34 in the smaller aperture
(within the central $r<5''\times2''$ or $350 \times 150$ pc region),
indicating that the nuclear HCN emission is compact
compared with the CO emission associated with the nucleus.
This is similar to the case of the type-1.5 Seyfert galaxy NGC 3227,
where unresolved HCN emission has been detected toward the nucleus
although the CO emission is extended across the nuclear region
(\cite{sch00}).

The observed CO and HCN distributions in NGC 1097 show 
a striking similarity to those in the type-2 Seyfert
galaxies NGC 1068 and NGC 5194, where extreme enhancements of $R_{\rm HCN/CO}$
have been reported (\cite{jac93, tac94, hb95, kk96}),
though many other Seyferts show no such significant enhancement of
$R_{\rm HCN/CO}$ values even with the high resolution ($\leq$ a few 100 pc)
observations (Kohno et al.\ 1999a, 1999b; \cite{we00}; \cite{hut00}).

What is the nature of this enhanced HCN emission toward some Seyfert nuclei?
It is tempting to speculate that a massive starburst occurs
in the very nucleus of ``the HCN enhanced Seyferts'' 
(i.e., NGC 1068, NGC 5194, and NGC 1097),
if we consider the quantitative
and spatial correlations between dense molecular gas
and massive star-forming regions in star-forming/starburst galaxies 
(e.g., \cite{sol92}; Kohno et al.\ 1999a). 
In fact, massive starbursts within the obscuring molecular torus
have been proposed (Cid Fernandes, Terlevich 1995),
and such intimate cohabitation of powerful
nuclear starburst and Seyfert activity has often been claimed
in the nuclei of some type-2 Seyferts
(e.g., Heckman et al.\ 1997; Gonz\'alez Delgado 1998; Maiolino et al.\ 1998).
Nevertheless, it seems to be too early to jump to that conclusion
because the reported high $R_{\rm HCN/CO}$ in NGC 1068, NGC 5194, and NGC 1097
(0.34 -- 0.6) are too high
compared with the centers of nearby nuclear starburst galaxies.
High resolution observations of local starbursts often show
the $R_{\rm HCN/CO}$ values within the range of about 0.1 to 0.2
(\cite{dow92}; \cite{pag95}; \cite{hb97}).
These ratios in starburst nuclei are indeed high compared with 
normal/quiescent galaxies (e.g., \cite{kk02}),
yet $R_{\rm HCN/CO}$ values exceeding 0.3 are quite rare (e.g., Sorai et al.\ 2002).

In the very centers of AGNs, the HCN abundance can be
enhanced due to strong X-ray emission from the active nuclei
(\cite{ld96}), and this could be related to the observed
enhancement of HCN emission in some Seyfert galaxies (\cite{tac94}).
Our preliminary results of the HCN/HCO$^+$ survey in Seyfert
and starburst galaxies imply that the enhanced HCN emissions
in the nuclei of NGC 1068, NGC 1097, and NGC 5194 could originate
from the X-ray irradiated molecular torii, and they are {\it not}
associated with the nuclear starbursts because the HCN/HCO$^+$ ratios
in NGC 1068, NGC 5194, and NGC 1097 are abnormally high
compared with nuclear starburst galaxies (\cite{kk01}).

\subsection{Dense Molecular Gas in the Circumnuclear Starburst Ring of NGC 1097}

We detected the CO and HCN emissions associated with 
the circumnuclear starburst ring in NGC 1097.
The azimuthally averaged $R_{\rm HCN/CO}$ is about 0.16 at the radius of the ring.
This ratio is comparable to those in the circumnuclear starburst rings
of the type-2 Seyfert galaxies NGC 1068 (\cite{hb95}) and NGC 6951 (Kohno et al.\ 1999a).
Most of the CO and HCN emission is located in the points where
offset dust lanes meet the starburst ring.
This situation is widely observed in the circumnuclear region 
of barred spiral galaxies, and is often referred as a ``twin-peaks'' morphology 
(\cite{ken92}).

Interestingly, there is a spatial shift between
CO and HCN twin-peaks; 
HCN peaks are shifted downstream compared with the CO peaks
if we assume trailing spiral arms (i.e., the gas rotates clockwise).
This spatial shift between CO and HCN in the circumnuclear ring
of barred spiral galaxies are reported in NGC 1530 (non-Seyfert: \cite{rd97})
and NGC 6951 (type-2 Seyfert: Kohno et al.\ 1999a).
Our CO data suggest the rotation velocity on the plane of the disk
is about 330 km s$^{-1}$ at a radius of $10''$ or 700 pc,
giving the dynamical time scale $T_{\rm dyn} \sim 1.3 \times 10^7$ yr.
Thus the observed spatial offset between the CO and HCN peaks,
about $20^\circ$ in the azimuth angle,
corresponds to a time delay of 
$\sim$ $(20/360)T_{\rm dyn} = 7.3 \times 10^5$ yr.
This time delay could be related to the time scale of dense molecular gas
formation due to gravitational instabilities of the molecular gas
(Kohno et al.\ 1999a).

These results may imply that the same mechanism can govern the star formation 
in the central $\sim$ kpc regions of both the type-1 and type-2 Seyfert galaxies.
It is often claimed that the host galaxies of the type-2 Seyfert nuclei 
harbor more intense star formation/starburst
compared with the type-1 Seyfert hosts (e.g., \cite{mai95,mai97}). 
Because the amount of the molecular gas in the type-2 Seyfert hosts 
seems to be similar to those in the type-1 hosts (\cite{mai97,vil98}), 
some mechanisms may exists to enhance the star formation efficiency
in the type-2 hosts if the putative enhancement of star formation 
in the type-2 Seyfert galaxies exists.
However, our high resolution CO and HCN images 
of the type-1 Seyfert galaxy NGC 1097 show striking similarities 
in the molecular gas distributions and physical properties of the molecular gas
in the central kpc regions of both the type-1 and type-2 Seyfert galaxies, 
though the sample is very limited at this moment. 
Further surveys of the high resolution molecular lines in type-1 and type-2
Seyferts could provide us with crucial keys 
whether significant differences on molecular gas properties
do exists or not between the Seyfert types.

\bigskip

The authors thank to the anonymous referee for the comments, which improve
this paper.
We acknowledge the support of the NRO staff for the operation of the telescopes
and the continuous efforts in improving the performance of the instruments.
Nobeyama Radio Observatory (NRO) is a branch of the National Astronomical Observatory,
an inter-university research institute operated by the Ministry of Education,
Culture, Sports, Sciene and Technology, Japan.


\begin{figure}
  \begin{center}
    \FigureFile(110mm,110mm){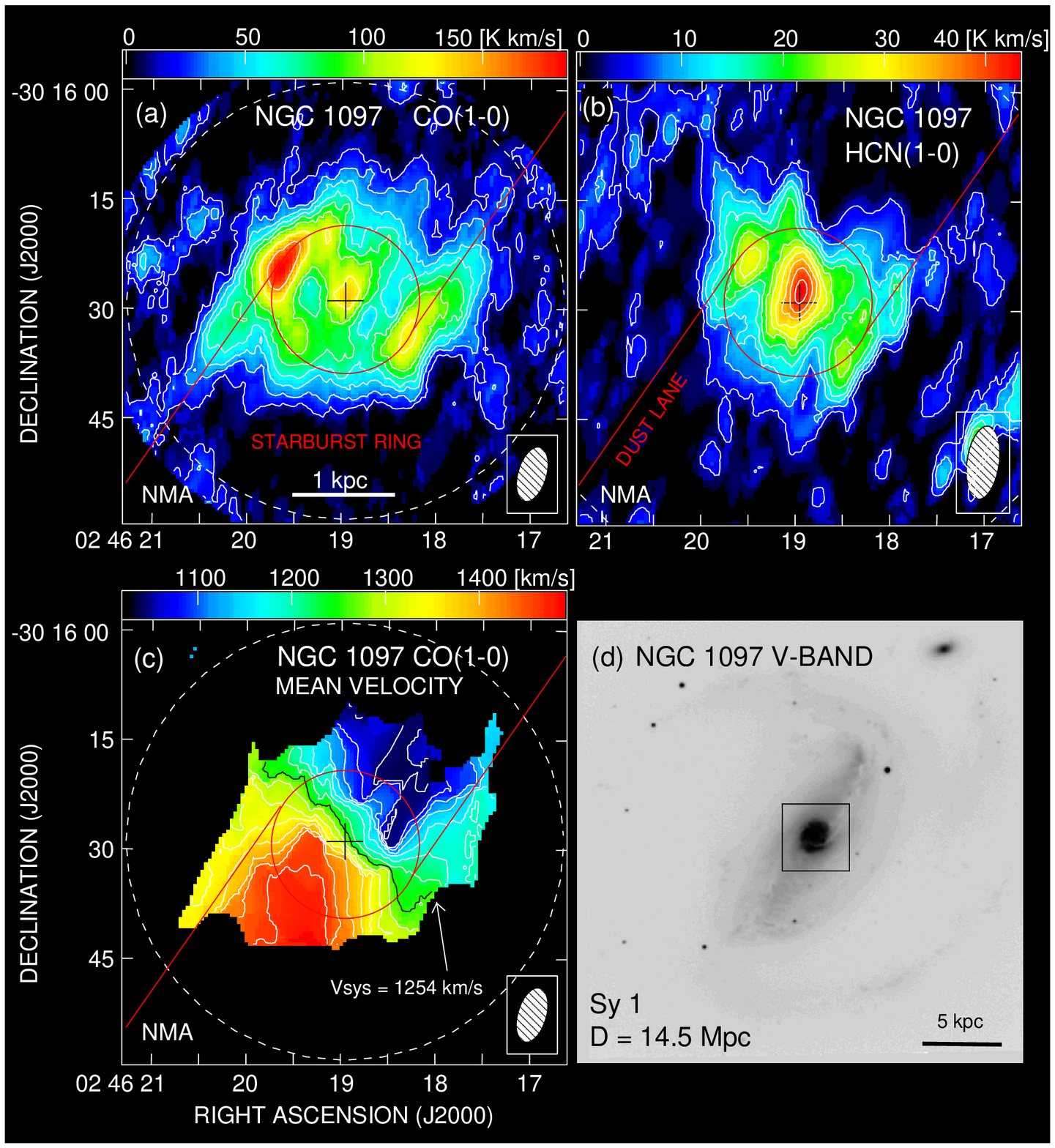}
  \end{center}
\caption{
The NMA maps of CO(1--0) and HCN(1--0) lines,
along with the optical image of NGC 1097.
The central cross in figure 1a, 1b, and 1c marks
the peak of 6 cm continuum (Hummel et al.\ 1987).
The coordinate is 
$\alpha$(J2000) = $02^{\rm h}46^{\rm m}18^{\rm s}\hspace{-5pt}.\hspace{2pt}96$ 
and $\delta$(J2000) = $-30^{\circ}16'28.\hspace{-2pt}''9$.
In each molecular map, the NMA field of view 
($60''$ for CO and $80''$ for HCN)
is indicated by a dashed circle. 
The circumnuclear starburst ring
with a radius of $r \sim 10''$ or 700 pc
is shown by a red circle,
and two red lines indicate the dust lanes along the bar.
Attenuation due to primary beam pattern of each 10 m dish
has been corrected in these maps.
(a) Integrated intensity map of CO.
The synthesized beam is \timeform{7".7} $\times$ \timeform{3".9}
($540 \times 270$ pc) with a P.A. of $-16^{\circ}$.
The contour levels are 1.5, 3, 4.5, 6, 7.5, 9, 12, 15, and 18 $\sigma$, 
where 1 $\sigma$ = 3.39 Jy beam$^{-1}$ km s$^{-1}$
or 10.4 K km s$^{-1}$ in $T_{\rm b}$. This corresponds to a face-on
gas surface density $\Sigma_{\rm gas}$ of 20.8 $M_\odot$ pc$^{-2}$,
calculated as $\Sigma_{\rm gas} = 1.36 \times \Sigma_{\rm H_2}$
and $\Sigma_{\rm H_2} = 2.89 \mbox{cos}\ i \times (I_{\rm CO}/\mbox{K km s$^{-1}$})$,
where $I_{\rm CO}$ is the velocity-integrated CO intensity
and $i$ is the inclination of the disk ($46^\circ$ for NGC 1097;
Ondrechen et al.\ 1989).
%
(b) Integrated intensity map of HCN.
The synthesized beam is \timeform{10".1} $\times$ \timeform{4".4}
($710 \times 310$ pc) with a P.A. of $-8^{\circ}$.
The contour levels are 1.5, 3, 4.5, $\cdots$, 15, and 16.5 $\sigma$,
where 1 $\sigma$ = 0.711 Jy beam$^{-1}$ km s$^{-1}$
or 3.54 K km s$^{-1}$ in $T_{\rm b}$. 
%
(c) Intensity-weighted mean velocity map of CO.
The contour interval is 30 km s$^{-1}$, and the systemic velocity of
1524 km s$^{-1}$, determined by the fitting of this CO data, 
is indicated.
Circular rotation seems to dominate the kinematics in the central
$r<10''$ region (i.e., within the circumnuclear ring) in NGC 1097, 
whereas very strong non-circular motions
along the dust lanes
are suggested by isovelocity contours nearly parallel to the dust lanes.
(d) A V-band image of NGC 1097 (Quillen et al.\ 1995). 
The central $400'' \times 400''$ (28 kpc $\times$ 28 kpc) 
region is displayed.
A solid box indicates the area of the CO and HCN maps
($60'' \times 60''$).
}\label{fig:NMAmaps}
\end{figure}

\begin{figure}
  \begin{center}
    \FigureFile(85mm,100mm){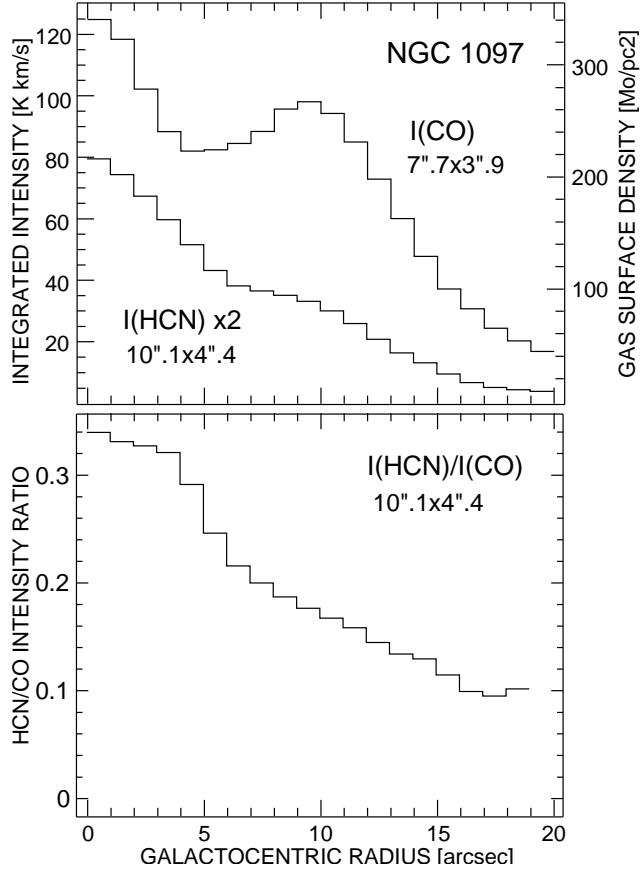}
  \end{center}
\caption{
Azimuthally averaged radial distributions of CO and HCN (top right),
and the HCN/CO ratios (bottom right) in NGC 1097. 
The systematic errors of the radial distributions
and the ratio are about $\pm$ 10 \%.
The face-on gas surface densities from the CO intensities
are also indicated,
assuming $X_{\rm CO}$ of $1.8 \times 10^{20}$ cm$^{-2}$ (K km s$^{-1}$)$^{-1}$.
}\label{fig:radialdist}
\end{figure}

\begin{figure}
  \begin{center}
    \FigureFile(85mm,100mm){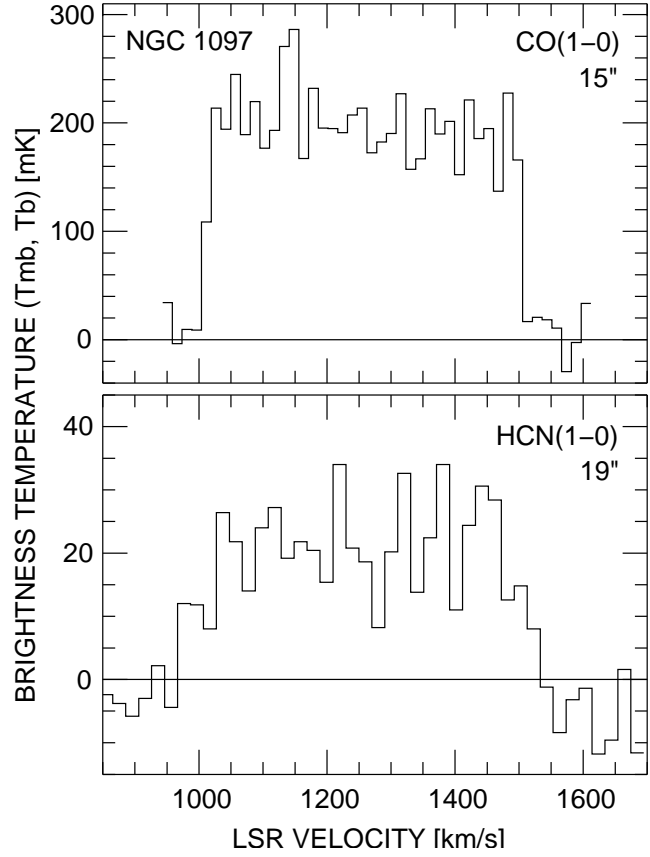}
  \end{center}
\caption{
CO and HCN spectra in the center of NGC 1097
obtained with the NRO 45 m telescope, 
displayed in the main-beam temperature scale 
($T_{\rm MB} \equiv T_{\rm A}^*/\eta_{\rm MB}$).
The beam sizes (HPBW)
of CO and HCN observations are $15''$ and $19''$, respectively.
The main-beam efficiencies ($\eta_{\rm MB}$) were 
$0.45 \pm 0.03$ for CO and
$0.50 \pm 0.03$ for HCN observations.
}\label{fig:45spectra}
\end{figure}


\begin{thebibliography}{}

\bibitem[Antonucchi(1993)]{ant93}
Antonucchi,~R. 1993, \araa, 31, 473

\bibitem[Barth et al.(1995)]{bar95} 
Barth,~A.~J., Ho,~L.~C., Filippenko,~A.~V., \& Sargent,~W.~L.~W.
1995, \aj, 110, 1009

\bibitem[Cid Fernandes,Terlevich(1995)]{cft95}
Cid Fernandes,~R., \& Terlevich,~R.  1995, \mnras, 272, 423

\bibitem[Curran et al.(2000)]{cur00}
Curran,~S.~J., Aalto,~S., \& Booth,~R.~S.  2000, \aaps, 141, 193

\bibitem[Dame et al.(2001)]{dame01}
Dame,~T.~M., Hartmann,~D., \& Thaddeus,~P. 2001, \apj, 547, 792

\bibitem[Downes et al.(1992)]{dow92}
Downes,~D., Radford,~S.~J.~E., Guilloteau,~S., Guelin,~M.,
Greve,~A., \& Morris,~D.  1992, \aap, 262, 424

\bibitem[Gerin et al.(1988)]{gnc88} 
Gerin,~M., Nakai,~N., \& Combes,~F. 1988, \aap, 203, 44

\bibitem[Helfer, Blitz(1993)]{hb93} 
Helfer~T., \& Blitz~L. 1993, \apj, 419, 86

\bibitem[Helfer, Blitz(1995)]{hb95} 
Helfer~T., \& Blitz~L. 1995, \apj, 450, 90

\bibitem[Helfer, Blitz(1997)]{hb97} 
Helfer~T., \& Blitz~L. 1997, \apj, 478, 162

\bibitem[Hummel et al.(1987)]{hum87} 
Hummel,~E., van der Hulst,~J.~M., Keel,~W.~C. 1987, \aap, 172, 32

\bibitem[H\"uttemeister et al.(2000)]{hut00}
H\"uttemeister,~S., Aalto,~S., Das,~M., \& Wall,~W.~F. 
2000, \aap, 363, 93

\bibitem[Jackson et al.(1993)]{jac93} 
Jackson,~J.~M., Paglione,~T.~D., 
Ishizuki,~S., \& Rieu,~N.~Q. 1993, \apj, 418, L13

\bibitem[Kenney et al.(1992)]{ken92} 
Kenney,~J.~D.~P., Wilson,~C.~D., Scoville,~N.~Z., 
Devereux,~N.~A., \& Young,~J.~S. 1992, \apj, 395, L79

\bibitem[Kohno et al.(1996)]{kk96} 
Kohno,~K., Kawabe,~R., Tosaki,~T., \& Okumura,~S.~K. 1996, \apj, 461, L29

\bibitem[Kohno et al.(1999a)]{kk99a} 
Kohno,~K., Kawabe,~R., \& Vila-Vilar\'o,~B. 1999a, \apj, 511, 157

\bibitem[Kohno et al.(1999b)]{kk99b} 
Kohno,~K., Kawabe,~R., \& Vila-Vilar\'o,~B. 1999b, 
in The Physics and Chemistry of the Interstellar Medium,
ed.\ V.~OssenKopf, J.~Stutzki, \& G.~Winnewisser (GCA-Verlag, Herdecke),
34 (astro-ph/9902251)

\bibitem[Kohno et al.(2001)]{kk01}
Kohno,~K., Matsushita,~S., Vila-Vilar\'o,~B., Okumura,~S.~K.,
Shibatsuka,~T., Okiura,~M., Ishizuki,~S., \& Kawabe,~R. 2001,
in The Central Kiloparsec of Starbursts and AGN: The La Palma
Connection, ed.\ J.~H.~Knapen, J.~E.~Beckman, I.~Shlosman, \&
T.~J.~Mahoney (ASP, San Francisco), 672 (astro-ph/0206398)

\bibitem[Kohno et al.(2002)]{kk02}
Kohno,~K., Tosaki,~T., Matsushita,~S., Vila-Vilar\'o,~B.,
\& Kawabe,~R.  2002, \pasj, 54, 541


\bibitem[Kotilainen et al.(2000)]{kot00} 
Kotilainen,~J.~K., Reunanen,~J., Laine,~S., \& Ryder,~S.~D.
2000, \aap, 353, 834

\bibitem[Lepp, Dalgarno(1996)]{ld96} 
Lepp,~S., \& Dalgarno,~A.  1996, \aap, 306, L21

\bibitem[Maiolino, Rieke(1995)]{mai95}
Maiolino,~R., \& Rieke,~G.~H.  1995,
\apj, 454, 95

\bibitem[Maiolino et al.(1997)]{mai97}
Maiolino,~R., Ruiz,~M., Rieke,~G.~H.,
\& Papadopoulos,~P.  1997, \apj, 485, 552

\bibitem[Iyomoto et al.(1996)]{iyo96} 
Iyomoto,~N., Makishima,~K., Fukazawa,~Y., Tashiro,~M., 
Ishisaki,~Y., Nakai,~N., \& Taniguchi,~Y.  1996, \pasj, 48, 231

\bibitem[Okumura et al.(2000)]{oku00}
Okumura,~S.~K., Momose,~M., Kawaguchi,~N., Kanzawa,~T., Tsutsumi,~T.,
Tanaka,~A., Ichikawa,~T., Suzuki,~T., \etal\ 
2000, \pasj, 52, 393

\bibitem[Ondrechen et al.(1989)]{ond89} 
Ondrechen,~M.~P., van der Hulst,~J.~M., \& Hummel,~E.
1989, \apj, 342, 39

\bibitem[Paglione et al.(1995)]{pag95} 
Paglione,~T.~A.~D., Tosaki,~T., \& Jackson,~J.M. 1995, \apj, 454 L117

\bibitem[Papadopoulos, Seaquist(1998)]{ps98}
Papadopoulos,~P.~P., \& Seaquist,~E.~R.  1998, \apj, 492, 521


\bibitem[P\'erez-Olea, Colina(1996)]{pc96} 
P\'erez-Olea D.E., \& Colina,~L. 1996, \apj, 468, 191

\bibitem[Quillen et al.(1995)]{qui95} 
Quillen,~A.~C., Frogel,~J.~A., Kuchinsky,~L.~E., \& Terndrup,~D.~M.
1995, \aj, 110, 156

\bibitem[Reynaud, Downes(1997)]{rd97}
Reynaud,~D., \& Downes,~D. 1997, \aap, 319, 737



\bibitem[Schinnerer et al.(2000)]{sch00}
Schinnerer,~E., Eckart,~A., \& Tacconi,~L.~J.  2000, \apj, 533, 826




\bibitem[Shaw et al.(1993)]{sha93}
Shaw,~M.~A., Combes,~F., Axon,~D.~J., \&  Wright,~G.~S.
1993, \aap, 273, 31


\bibitem[Solomon et al.(1992)]{sol92} 
Solomon,~P.~M., Downes,~D., \& Radford,~S.~J.~E. 
1992, \apj, 387, L55


\bibitem[Sorai et al.(2002)]{sol02} 
Sorai,~K., Nakai,~N., Kuno,~N., \& Nishiyama,~K.
2002, \pasj, 54, 179 

\bibitem[Storchi-Bergmann et al.(1997)]{sto97}
Storchi-Bergmann,~T., Eracleous,~M., Ruiz,~M.~T., Livio,~M.,
Wilson,~A.~S., \& Filippenko,~A.~V.  1997, \apj, 489, 87



\bibitem[Tacconi et al.(1994)]{tac94} 
Tacconi,~L.~J., Genzel,~R., Blietz,~M., Cameron,~M., 
Harris,~A.~I., \& Madden,~S.  1994, \apj, 426, L77


\bibitem[Telesco et al.(1993)]{tel93}
Telesco,~C.~M., Dressel,~L.~L., \& Wolstencroft,~R.~D. 
1993, \apj, 414, 120


\bibitem[Tully(1988)]{tul88} 
Tully,~R. 1988, Nearby Galaxies Catalog 
(Cambridge University Press, Cambridge)

\bibitem[Vila-Vilar\'o et al.(1998)]{vil98}
Vila-Vilar\'o,~B., Taniguchi,~Y., \& Nakai,~N.\ 1998, \aj, 116, 1553


\bibitem[Wild, Eckart(2000)]{we00}
Wild,~W., \& Eckart,~A.  2000, \aap, 359, 483


\bibitem[Wehrle et al.(1997)]{weh97}
Wehrle,~A.~E., Keel,~W.~C., \& Jones,~D.~L.  1997, \aj, 114, 115


\end{thebibliography}
\end{document}